\documentclass[prl,aps,superscriptaddress,showpacs,nofootinbib]{revtex4-1}
\usepackage{graphicx,graphics,epsfig}
\usepackage{amsmath}
\usepackage{verbatim}
\usepackage{color}
\usepackage[standard]{ntheorem}

\definecolor{myurlcolor}{rgb}{0,0,0.4}
\definecolor{mycitecolor}{rgb}{0,0.5,0}
\definecolor{myrefcolor}{rgb}{0.5,0,0}
\usepackage[pagebackref]{hyperref}
\hypersetup{colorlinks, linkcolor=myrefcolor,
citecolor=mycitecolor, urlcolor=myurlcolor}

\newcommand{\id}{\openone}
\newcommand{\ket}[1]{|#1\rangle}

\definecolor{nblue}{rgb}{0.2,0.2,0.8}
\definecolor{ngreen}{rgb}{0.2,0.8,0.2}
\definecolor{nred}{rgb}{0.8,0.2,0.2}
\definecolor{nblack}{rgb}{0,0,0}

\date{\today}

\begin{document}

\title{Demonstrating quantum contextuality of indistinguishable particles by a single family of noncontextuality inequalities}

\author{Hong-Yi Su\footnote{Correspondence to
H.Y.S. (hysu@mail.nankai.edu.cn).}}
 \affiliation{Theoretical Physics Division, Chern Institute of Mathematics, Nankai University,
 Tianjin 300071, People's Republic of China}

\author{Jing-Ling Chen\footnote{Correspondence to
J.L.C. (chenjl@nankai.edu.cn).}}
 \affiliation{Theoretical Physics Division, Chern Institute of Mathematics, Nankai University,
 Tianjin 300071, People's Republic of China}
 \affiliation{Centre for Quantum Technologies, National University of Singapore,
 3 Science Drive 2, Singapore 117543}

\author{Yeong-Cherng Liang\footnote{Correspondence to
Y.C.L. (yliang@phys.ethz.ch).}}
 \affiliation{Institute for Theoretical Physics, ETH Zurich, 8093 Zurich, Switzerland}

\date{\today}


\maketitle

\textbf{Quantum theory has the intriguing feature that is inconsistent with noncontextual hidden variable models, for which the outcome of a measurement does not depend on which other compatible measurements are being performed concurrently. While various proofs of such contextual behavior of quantum systems have been established, relatively little is known concerning the possibility to demonstrate this intriguing feature for indistinguishable particles.
Here, we show in a simple and systematic manner that with projective measurements alone, it is possible to demonstrate quantum contextuality for such systems of arbitrary Hilbert space dimensions, including those corresponding to a qubit. Our demonstration is applicable to a single fermion as well as multiple fermions, and thus also a composite boson formed from an even number of fermions.
In addition, our approach gives a clear demonstration of the  intimate connection between complementarity and contextuality, two seemingly unrelated aspects of quantum theory.
}

A fundamental feature of quantum theory is that measurement outcomes generally cannot be predicted with certainty even with precise knowledge of the measurement procedure as well as the state of the system. Is this randomness  unavoidable or could there be some higher-level theory that ``completes"~\cite{EPR} quantum theory and restores determinism by supplementing the latter with additional hidden variables (HV)? Bohmian mechanics~\cite{Bohm1,Bohm2} is one such example. Could there be others?
In the 1960s, Bell~\cite{Bell:RMP} and independently Kochen \& Specker~\cite{KS} showed that quantum theory is incompatible with  the assumption underlying the so-called (measurement-outcome-) noncontextual (NC) HV theory~\cite{Peres,Spekkens:PRA:71}. Loosely, such theories assume that the measurement outcome is independent of the measurement contexts. For instance, if \textbf{A} and \textbf{B} are compatible measurements, likewise for \textbf{A} and \textbf{C}, such a theory demands that the measurement outcome of \textbf{A} is independent of whether \textbf{A} is measured together with \textbf{B} or with \textbf{C}.

The aforementioned incompatibility is now commonly referred to as quantum contextuality (QC). This nonclassical feature of quantum theory, in particular a strong form of it known as Bell-nonlocality~\cite{Bell64,Brunner:RMP}, has triggered a lot of discussions about some of the very fundamental concepts in physics that we have taken for granted (see, eg., Refs.~\cite{Liang, Peres, Foundations1,Foundations2,Foundations3,Foundations4,Foundations5,Foundations6,Foundations7,Foundations8,Spekkens:1312.3667,Foundations10} and references therein). On the more pragmatic side, it is worth noting that Bell-nonlocality is known as an indispensable resource in device-independent quantum information processing~\cite{Brunner:RMP,Scarani:DIQIP}, whereas QC itself has recently been argued to be {\em the} resource~\cite{Howard:Nature:2014} that enables quantum computing~\cite{Nielsen}.

Traditional proofs of contextuality, such as the one given by Kochen and Specker~\cite{KS}, though elegant and rigorous, are not without their drawbacks. Firstly, such proofs rely heavily on the structure of Hilbert space and thus lack a clear operational meaning --- an essential feature required for any meaningful experimental test against another operational theory. Secondly, they are only applicable to projective measurements, thus rendering a proof of contextuality impossible in any two-dimensional Hilbert space. In recent years, the first drawback has been overcome to some extent by an approach pioneered by Klyachko {\em et al.}~\cite{KCBS} based on noncontextuality inequalities --- constraints on measurement statistics  necessarily satisfied by any measurement-outcome NCHV theory (see also Refs.~\cite{Cabello:1998,Simon:PRL,Larsson,Spekkens:PRA:71} for other operationally-inspired works in this regard). The approach of Klyachko {\em et al.} was further developed, e.g., in Refs.~\cite{Liang,Cabello08,Cabello09,spin-1/2,NCI,Araujo} and led to a series of experiments verifying quantum contextuality~\cite{KCBS-exp,Blatt09,exp1,exp2,exp3} (for loopholes that could still apply to such operationally-based tests, see, e.g., Refs.~\cite{Meyer,Kent,KentClifton,Winter:JPA}).

Concerning the inapplicability of traditional proofs of contextuality in a two-dimensional Hilbert space, some possible workarounds~\cite{Cabello:PRL:90,Busch:PRL} have been proposed using generalized (unsharp) measurements described by positive-operator-valued measure (POVM)~\cite{Nielsen}. These works, however, assumed  deterministic outcomes even for POVM, an assumption that is debatable, as remarked in Refs.~\cite{Spekkens:PRA:71,Liang,Spekkens:1312.3667} (see also Ref.~\cite{criticize} for other criticism). In turn,  by considering also only POVM, Spekkens provided~\cite{Spekkens:PRA:71} alternative proof of measurement-outcome contextuality, as well as other operationally-motivated notions of quantum contextuality for two-dimensional quantum systems. In this work, as a consequence of our demonstration of quantum contextuality for indistinguishable particles via noncontextuality inequalities, we provide a different workaround to the aforementioned problem --- a demonstration of qubit contextuality using {\em projective measurements} and the physical constraints that stem from the (anti-)commutation relations of  indistinguishable particles. Note that the contextuality of systems of indistinguishable particles was also studied earlier in Ref.~\cite{identical} by considering all degrees of freedom associated with such systems (see also Ref.~\cite{spin-1/2}). Here, we show that for the demonstration of their contextual behavior, it suffices to consider  their discretized momentum (or position) degree of freedom.

The rest of this paper is organized as follows: we first consider a fermionic wavepacket of two momenta and present a noncontextuality inequality
to manifest its QC. Then we generalize the situation to a wavepacket involving an arbitrary number of momentum modes. For a planewave with a definite momentum, we apply the noncontextuality inequality to its complementary degree of freedom to demonstrate its QC.
The generalization to the case of more than one fermion is demonstrated, before discussing the fermion-fermion case, which proves the contextual behavior also of a composite boson. We will discuss the close relation between quantum complementarity and QC towards the end of this article.

\vspace{5mm}
\noindent{\bf Results}

\noindent{\bf Preliminaries associated with fermions.} Let us begin by reminding that it is convenient to use the Fock basis representation to describe the (anti)symmetric states of indistinguishable particles. In general, a fermionic one-particle state in a {\em finite volume} can be expressed as~\cite{single}
\begin{eqnarray}
|\psi\rangle=\sum_{j=1}^M g_{(k_j)} a_{k_j}^\dagger|\Omega\rangle,\label{psi}
\end{eqnarray}
where
$g_{(k_j)}$'s are complex coefficient satisfying $\sum_j |g_{(k_j)}|^2=1$, $|\Omega\rangle$ denotes the vacuum state, $a_{k_j}^\dagger$ (and its adjoint $a_{k_j}$) is the creation (annihilation) operator for momentum mode $k_j$ satisfying the equal-time anticommutation relations
\begin{eqnarray}
\{a_{k_j},a_{k_{j'}}\}=0,\;\;\;\;\;
\{a_{k_j},a_{k_{j'}}^\dagger\}=\delta_{k_j k_{j'}},
\end{eqnarray}
and $\delta$ is the Kronecker delta function. Define the number operator $\hat N=\sum_{j=1}^M a_{k_j}^\dagger a_{k_j}$,
then it is obvious that $\langle\psi|\hat N|\psi\rangle=1$, indicating that Eq.~(\ref{psi}) represents a one-particle state, as claimed.

If the volume is infinite, one must make the substitutions $\delta_{k_j k_{j'}}\rightarrow\delta(k-k')$ and $\sum_j\rightarrow\int{\rm d}k$. However, it is expedient to consider first the finite case and then set the volume arbitrarily large to effectively approximate the continuous case. In fact, most physically continuous quantities are discrete in experiments, since the laboratory itself is to be understood as a finite box. As another example, the continuous energy band in solid-state physics is formed by combining a large number of atoms which possess a discrete set of energy levels.

Without loss of generality, consider that only $M$ fermionic modes have non-zero occupancy with momenta $k_j$ in increasing order, i.e.,
\begin{eqnarray}
k_1<k_2<\cdots<k_M.
\end{eqnarray}
To reveal the quantum contextuality of a single fermion, we shall --- motivated by the Pauli exclusion principle $a_{k_j}^2=a_{k_j}^{\dagger\,2}=0$ --- construct Pauli observable $\sigma$ from the creation (annihilation) operators for each mode.
To this end, we employ the Jordan-Wigner transformation~\cite{JW-trans} in the ``reciprocal" space:
\begin{eqnarray}
\sigma_{k_j}^+&=&\exp\biggr({-i\pi\sum_{m=0}^{j-1}a_{k_m}^\dagger a_{k_m}}\biggr)a^\dagger_{k_j},\label{sigma+}\\
\sigma_{k_j}^-&=&\exp\biggr({i\pi\sum_{m=0}^{j-1}a_{k_m}^\dagger a_{k_m}}\biggr)a_{k_j},\label{sigma-}\\
\sigma_{k_j}^z&=&2a_{k_j}^\dagger a_{k_j}-\openone,\label{sigmaz}
\end{eqnarray}
where $\openone$ is the identity operator and we define $a_{k_0}^\dagger a_{k_0}\equiv0$ for consistency. One can now readily verify that
\begin{eqnarray}
[\sigma^+_{k_j},\sigma^-_{k_{j'}}]=\delta_{k_j k_{j'}}\sigma^z_{k_j},\;\;\;
[\sigma^z_{k_j},\sigma^{\pm}_{k_{j'}}]=\pm\delta_{k_j k_{j'}}\sigma^{\pm}_{k_j}.
\end{eqnarray}
If we further define $\sigma^x_{k_j}$ and $\sigma^y_{k_j}$ via $\sigma^\pm_{k_j}=\frac{1}{2}(\sigma^x_{k_j}\pm i\sigma^y_{k_j})$,
the desired observable can then be written as:
\begin{eqnarray}
&&\sigma_{k_j}^{\hat n_j}\equiv\vec\sigma_{k_j}\cdot\hat n_j\label{sigma}\\
&&=\biggr(\prod_{m=0}^{j-1}(\id-2a_{k_m}^\dagger a_{k_m})\biggr)\biggr(e^{i\phi_j}a_{k_j}+e^{-i\phi_j}a_{k_j}^\dagger\biggr)\sin\theta_j\nonumber\\
&&\;\;\;\;\;\;\;\;\;\;\;\;\;\;\;\;\;\;\;\;\;\;\;\;\;\;\;\;\;\;\;\;\;\;\;\;\;\;\;\;\;\;\;\;\;\;\;\;
+(2a_{k_j}^\dagger a_{k_j}-\id)\cos\theta_j,\nonumber
\end{eqnarray}
where $\vec\sigma_{k_j}=(\sigma_{k_j}^x,\sigma_{k_j}^y,\sigma_{k_j}^z)$ is the vector of Pauli matrices defined through Eqs.~(\ref{sigma+}), (\ref{sigma-}) and (\ref{sigmaz}), and $\hat n_j=(\sin\theta_j\cos\phi_j,\sin\theta_j\sin\phi_j,\cos\theta_j)$ is a unit vector in $\mathbb{R}^3$.
Note that the first product term in Eq.~(\ref{sigma}) results from the exponentials in Eqs.~(\ref{sigma+}) and (\ref{sigma-}).

As a fermionic realization of the Pauli operator, here $\sigma_{k_j}^{\hat n_j}$ shares familiar properties: it takes eigenvalues $\pm 1$ and has corresponding eigenprojector
$P(\pm|{k_j})=[1\pm(-1)\sigma_{k_j}^{\hat n_j}]/2$. We show in \textbf{Methods} that for a general fermionic state
\begin{equation}
	|\psi_{\rm F}\rangle=\sum_{\mu} t_{\mu}(a_{k_1}^\dagger)^{\mu_1}(a_{k_2}^\dagger)^{\mu_2}\cdots
	(a_{k_M}^	\dagger)^{\mu_M}|\Omega\rangle,\label{QFT2}
\end{equation}
with $\mu=(\mu_1,\mu_2,...\mu_M)$ being an $M$-tuple of binary-valued element $\mu_j=0,1$, there always exists a corresponding state $|\psi_{\rm H}\rangle=\sum_{\mu} t_{\mu}|\mu_1,\mu_2,\ldots,\mu_M\rangle$ in the Hilbert space ${\mathbb{C}^2}^{\otimes M}$ such that
\begin{eqnarray}
	&&\langle\psi_{\rm F}|\sigma_{k_1}^{\hat n_1}\sigma_{k_2}^{\hat n_2}\cdots \sigma_{k_M}^{\hat n_M}|\psi_{\rm F}\rangle		
	\nonumber\\
	&&\;\;\;\;\;\;\;\;\;\;\;\;\;\;\;=\langle\psi_{\rm H}|\hat\sigma_1^{\hat n_1}\otimes \hat\sigma_2^{\hat n_2}\otimes\cdots
	\otimes\hat\sigma_M^{\hat n_M}|\psi_{\rm H}\rangle,\label{correspond}
\end{eqnarray}
where the hatted operator $\hat\sigma_j^{\hat n_j}$ is defined as:
\begin{eqnarray}
\hat\sigma_j^{\hat n_j}=\left(\begin{matrix}-\cos\theta_j & \sin\theta_j e^{i\phi_j}\\ \sin\theta_j e^{-i\phi_j} & \cos\theta_j\end{matrix}\right).\label{sigma-QM}
\end{eqnarray}
It is worth noting that there is no direct relation between the $M$-qubit state $|\psi_{\rm H}\rangle$ and the number of particles in $|\psi_{\rm F}\rangle$, since the number of constituent Hilbert spaces that we need to define $|\psi_{\rm H}\rangle$ is equal to the number of distinct momentum modes $M$, rather than the number of particles (excitations) $N$. Obviously $M$ is lower bounded by $N$ for fermions, but they are, otherwise, independent quantities.

\vspace{5mm}

\noindent{\bf Contextuality of a fermion in two momentum modes.} Let us now demonstrate the contextual behavior with the fermionic state (\ref{psi}), cf. Eq.~\eqref{QFT2}, focusing first on the case where the momentum takes only two distinct values $k_1$ and $k_2$ (i.e., $M=2$). Consider now the CHSH Bell inequality~\cite{CHSH}, which can also be seen as a noncontextuality inequality,
\begin{align}
I(k)&=E(k_1 ,k_2)+E(k_1,k_2')+E(k_1',k_2)-E(k_1',k_2')\nonumber\\
&\stackrel{\mbox{\tiny NC}}{\leq}2\label{chsh}
\end{align}
where $E(k_i,k_j)$ is the expectation value corresponding to the joint measurement of the observables labeled, respectively, by $k_i$ and $k_j$, cf. Eq.~\eqref{sigma}; likewise, we use the symbol $k_j'$ as a label for an observable associated with the momentum mode $k_j$ but for the primed unit vector $\hat{n}_j'.$ (The symbol $k_j'$ is not to be confused with $k_{j'}$, which refers to a momentum mode different from $k_j$.) Note that the upper bound dictated by NCHV can be easily verified by considering deterministic measurement outcome for each of these measurements.

From Eq.~\eqref{sigma}, it can be shown that the commutators $[\sigma_{k_i}^{\hat n_i},\sigma_{k_j}^{\hat n_j}]$, $[\sigma_{k_i}^{\hat n_i},\sigma_{k_j}^{\hat n_j'}]$ vanish for $i\neq j$ and thus the observable corresponding to different values of $k_j$ are indeed jointly measurable. The expectation value of Eq.~\eqref{chsh} then takes the explicit form of $E(k_1,k_2)=\langle\sigma_{k_1}^{\hat n_1}\sigma_{k_2}^{\hat n_2}\rangle=\langle\psi_{\rm F}|\sigma_{k_1}^{\hat n_1}\sigma_{k_2}^{\hat n_2}|\psi_{\rm F}\rangle$, and similarly for the other terms.
In particular, by setting
\begin{eqnarray}
&\theta_1=\pi,\;\;\;\theta_2=\arctan[2g_{(k_1)}g_{(k_2)}],&\nonumber\\
&\theta'_1=\frac{\pi}{2},\;\;\;\theta'_2=-\theta_2,&\\
&\phi_1=\phi_2=0,\;\;\;\phi'_1=\phi'_2=0,&\nonumber
\end{eqnarray}
the left-hand-side of Eq.~\eqref{chsh} becomes
\begin{eqnarray}
I^{\rm max}(k)=
2\sqrt{1+4g_{(k_1)}^2 g_{(k_2)}^2},
\end{eqnarray}
giving the maximal quantum value of the CHSH expression for given $|\psi_F\rangle$. Thus, except when $g_{(k_j)}=0$ for some $k_j$, the single fermion state defined in Eq.~(\ref{psi}) is always incompatible with noncontextuality for $M=2$.

Three remarks are now in order.  Firstly, since $\sigma_{k_2}^{\hat n_2}$ also includes contribution(s) from $a_{k_1}$ and $a_{k_1}^\dagger$, $\sigma_{k_1}^{\hat n_1}\sigma_{k_2}^{\hat n_2}$ generally cannot be factorized into a product form such as $u(a_{k_1},a_{k_1}^\dagger)v(a_{k_2},a_{k_2}^\dagger)$.
Secondly, for the case of two distinct momenta modes $|k_1\rangle$ and $|k_2\rangle$ (i.e., $M=2$), the fermionic property implies that a general state of the fermion described by Eq.~\eqref{psi} only has support in a two-dimensional Hilbert space. In standard quantum information terminology, such a fermion therefore defines a {\em qubit}~\cite{Nielsen} through its momentum degree-of-freedom. Thus, the QC identified above applies essentially to all pure states of a single qubit.
(This does not contradict the known result that qubit contextuality cannot be established using projective measurements. We will come back to this subtle point towards the end of the article.)

Thirdly, let us note that the mapping  established in Eq.~\eqref{correspond} also implies an analogous correspondence between any mixed state describing a single fermion with $M=2$ and some mixed state in $\mathbb{C}^2\otimes\mathbb{C}^2$. Since all two-qubit mixed states are incompatible with noncontextuality~\cite{Cabello08}, any mixed state $\rho=\sum_i p_i |\psi^i\rangle\langle\psi^i|$ describing such fermionic system --- with $|\psi^i\rangle$ being mutually orthogonal states of the form of Eq.~\eqref{psi} --- are also incompatible with noncontextuality. For instance, using Eq.~\eqref{correspond} and a Peres-Mermin-square~\cite{Mermin, Mermin-Peres}-type construction, one can see that any fermionic mixed state with $M=2$ violates  the noncontextuality inequality~\cite{Cabello08}
\begin{eqnarray}
\langle \mathcal{O}_{11}\mathcal{O}_{12}\mathcal{O}_{13}\rangle+\langle
\mathcal{O}_{21}\mathcal{O}_{22}\mathcal{O}_{23}\rangle+\langle \mathcal{O}_{31}\mathcal{O}_{32}\mathcal{O}_{33}\rangle+\nonumber\\
\langle \mathcal{O}_{11}\mathcal{O}_{21}\mathcal{O}_{31}\rangle+\langle
\mathcal{O}_{12}\mathcal{O}_{22}\mathcal{O}_{32}\rangle-\langle \mathcal{O}_{13}\mathcal{O}_{23}\mathcal{O}_{33}\rangle\stackrel{\mbox{\tiny NC}}{\leq} 4,
\label{PM1}
\end{eqnarray}
where each $\mathcal{O}_{ij}\;(i,j=1,2,3)$ is a dichotomic observable corresponding to the $(i,j)$ entry in
\begin{eqnarray}
\mathcal{O}=\left(\begin{matrix} \sigma_{k_1}^{\hat n_1} & \sigma_{k_2}^{\hat n_2} & \sigma_{k_1}^{\hat n_1}\sigma_{k_2}^{\hat n_2} \\
\sigma_{k_2'}^{\hat n_2} & \sigma_{k_1'}^{\hat n_1} &\sigma_{k_1'}^{\hat n_1}\sigma_{k_2'}^{\hat n_2}\\
\sigma_{k_1}^{\hat n_1}\sigma_{k_2'}^{\hat n_2} & \sigma_{k_1'}^{\hat n_1}\sigma_{k_2}^{\hat n_2} & \sigma_{k_1''}^{\hat n_1}\sigma_{k_2''}^{\hat n_2}
\end{matrix}\right),\label{PM}
\end{eqnarray}
with $\hat n_1=\hat n_2=(0,0,1)$, $\hat n_1'=\hat n_2'=(1,0,0)$, $\hat n_1''=\hat n_2''=(0,1,0)$.

\vspace{5mm}

\noindent{\bf Contextuality of a fermion in an arbitrary number of momentum modes.}  Let us now consider the case of $M$ distinct momentum modes. In analogy with the previous case, such a fermion therefore defines a qudit (with $d=M$) via its momentum degree-of-freedom.
As with the $M=2$ case, the correspondence of Eq.~\eqref{correspond} allows us to map any pure fermionic state $\ket{\psi_F}$ with $M\ge2$ to a pure state $\ket{\psi_H}$ in  ${\mathbb{C}^2}^{\otimes M}$. This, in turn, suggests that we can reveal the QC of $|\psi\rangle$ via existing multipartite Bell inequalities.
To this end, we consider the Hardy inequality~\cite{Hardy1,Hardy2}:
\begin{eqnarray}
I_{\rm Har}(k)&=&P(00\cdots 0|k_1 k_2\cdots k_M)\nonumber\\
&&-P(00\cdots0|k'_1 k_2\cdots k_M)\nonumber\\
&&-P(00\cdots0|k_1 k'_2\cdots k_M)\nonumber\\
&&-\cdots-P(00\cdots0|k_1 k_2\cdots k'_M)\nonumber\\
&&-P(11\cdots1|k'_1 k'_2\cdots k'_M)\stackrel{\mbox{\tiny NC}}{\leq} 0, \label{hardy}
\end{eqnarray}
where $P(\mu_1\mu_2\cdots\mu_M|k_1 k_2\cdots k_M)$ denotes the joint conditional probability of observing outcomes $\mu_1\mu_2\cdots\mu_M$ given measurements labeled by $k_1 k_2\cdots k_M$.
When $M=2$, inequality (\ref{hardy}) reduces to the CHSH inequality~(\ref{chsh}), up to permutations of ($k_1,k_2$), and of ($\mu_j,1-\mu_j$).

Using inequality (\ref{hardy}), it was shown that {\em all} pure entangled states violate Bell inequalities~\cite{Yu}. Recall from Eq.~(\ref{psi}) that in the Fock basis representation, $|\psi\rangle$ takes the form of a generalized $W$ state~\cite{W-state1},
which is typically a multipartite entangled state. As a result, we can see from the correspondence given in Eq.~\eqref{correspond} that the fermionic state (\ref{psi}) generically violates inequality (\ref{hardy}), showing QC for an arbitrary $M\ge2$. Hence, QC is a
ubiquitous feature demonstrated with states like (\ref{psi}) or any others with coherently distributed fermionic modes in Fock spaces. An important point to note now is that although we made use of the nonlocal nature with all pure entangled states shown in Ref.~\cite{Yu}, this result by itself does not demonstrate the contextuality with all pure quantum states ---  the correspondence that we have provided in Eq.~\eqref{correspond} is still needed to establish the missing link.

Obviously, the form of Eq.~(\ref{psi}) implies a mixture of different momenta. It thus seems like the demonstration of QC with such a fermionic state requires non-vanishing momentum uncertainty in the physical system. We now make use of the complementarity principle to argue that this is not the case.
Let us consider the noncontextuality inequalities~\eqref{chsh} and~\eqref{hardy}, but applied to the complementary degree of freedom, namely, via the substitution of $k\rightarrow x$,
\begin{eqnarray}
I(k)\rightarrow I(x)\leq2,\;\;\;{\rm and}\;\;\;
I_{\rm Har}(k)\rightarrow I_{\rm Har}(x)\leq0,\label{hardy2}
\end{eqnarray}
with $x$ being the position.
According to Heisenberg's uncertainty relation, for a state with definite momentum, the number of position modes $M'$ involved in Eq.~(\ref{hardy2}) must go to infinity, which is, however, ill-defined in mathematical rigor. Instead, let us assume that $M'$ is arbitrarily  large but not infinite.
In this way, we can once again demonstrate QC even if the fermion has a well-defined momentum.
Now, let
\begin{eqnarray}
\frac{1}{\mathcal{N}}\int{\rm d}x \;e^{ik_j x}|x\rangle\equiv|k_j\rangle\rightarrow|\psi\rangle=\frac{1}{\sqrt{L}}\sum_{m=1}^{ M'} e^{ik_j x_m}|x_m\rangle.\nonumber\\ \label{x}
\end{eqnarray}
The state of a fixed momentum is a superposition of all planewaves with the same $k_j$, and can be effectively approximated by fermionic state having discrete position modes. This is the essential idea of the quantum complementarity principle: one cannot learn the precise values of two mutually conjugate observables. By applying Eq.~(\ref{hardy2}), the QC of a fermion with a definite momentum can also be effectively detected via judicious choices of measurements. In this regard, it can be seen that the quantum complementarity principle plays a very interesting role in identifying QC: If it failed, the quantum violation of the pair of inequalities could both be zero, and henceforth no QC could be detected via our approach.

Analogous to the qubit case $M=2$, an alternative proof for the contexuality of a single fermion occupying $M\ge2$ momentum modes is also  possible by resorting to a generalized construction of Peres-Mermin square~\cite{planat12}. The advantage of such an alternative proof is that the resulting proof can be applied to an arbitrary mixed fermionic state with any $M\ge 2$.

\vspace{5mm}
\noindent{\bf Generalization to a scenario of multiple fermions.}
The above results can be generalized to the case of more than one fermion with no difficulty. In general, an $N$-fermion state (with $M\geq N$) is expressed as
\begin{eqnarray}
|\psi_N\rangle=
\sum_{j_1<j_2<\cdots<j_N\in[1,M]} g_{(k_{j_1},k_{j_2},\cdots,k_{j_N})} \;\;\;\;\;\;\;\label{psiN}\\ \times a_{k_{j_1}}^\dagger a_{k_{j_2}}^\dagger\cdots a_{k_{j_N}}^\dagger|\Omega\rangle,\nonumber
\end{eqnarray}
which, in the Fock basis representation, takes the form of a (generalized) Dicke state~\cite{Dicke}. Thus, unless $g_{(k_{j_1},k_{j_2},\cdots,k_{j_N})}$ is non-vanishing for only one term in the sum, the correspondence established in Eq.~\eqref{correspond} again maps $|\psi_N\rangle$ to an $M$-qubit entangled state which allows for a proof of QC in a similar manner.
The case of $N=2\ell$ with $\ell\ge1$ being a positive integer is of particular interest, since an even number of fermions constitute a composite boson. As a result, we are also able to identify the QC for such bosons. Hence, \emph{all species of indistinguishable particles can be incompatible with noncontextuality by the violation of a family of noncontextuality inequalities}. (Strictly, for the case where the $2\ell$ fermions are delocalized into $M=2\ell$ modes, we again need to invoke the complementarity between position and momentum as well as the finite approximation of infinitely many position modes.)

\vspace{5mm}
\noindent\textbf{Discussion}

\noindent To summarize, we have demonstrated the QC of a system of indistinguishable particles consisting of fermions, in particular a single fermion, through its quantum violations of a family of noncontextuality inequalities. The fermionic commutation relations play an essential role in our reasoning, so as the quantum complementarity principle.
Together, they guarantee the violation of  noncontextuality inequalities applied to at least one of the complementary degrees of freedom (such as position and momentum) of the system, thereby demonstrating QC of fermionic systems in general.
Obviously, a straightforward application of our result to composite bosons formed from, say, two fermions also demonstrate the QC of this other kind of indistinguishable particles. The possibility to extend our argument to a single elementary boson remains as an open problem.

Let us now come back to the apparent inconsistency between our result, which demonstrates the QC of a single fermion occupying an arbitrary number of momentum modes (including two),  and the well-known fact that in a two-dimensional Hilbert space, it is {\em impossible} to demonstrate QC by considering only (rank-1) projective measurements. This no-go theorem stems from the fact that in a two-dimensional Hilbert space, the ``context" of a projective measurement is fully determined by specifying any of its POVM elements. In our proof, although the {\em physical state} of the single particle (in the case of two momenta modes) is a qubit, the observables that we consider are ``mathematically" well defined even for the vacuum $|\Omega\rangle$ and the two-mode state $a_{k_1}^\dag a_{k_2}^\dag|\Omega\rangle$. Hence, our proof in some sense {\em does} make use of the mathematical structure of a higher-dimensional Hilbert space and does not contradict the well-known no-go result.
Note that unlike in quantum mechanics, here any superposition like $(a_{k_1}^\dag +a_{k_2}^\dag a_{k_3}^\dag)|\Omega\rangle$ is forbidden, due to the fermion-boson superselection rule. In other words, it is possible to prove that qubit is incompatible with noncontextuality by using projective measurements when supplemented with additional physical assumptions (e.g., the anti-commutation relations of fermions considered in this paper).

Given that we made use of complementarity in our proofs of contextuality, one may ask if complementarity is indeed a necessary ingredient (either implicitly or explicitly) for the proof of contextuality. In other words, does contextuality imply complementarity? The answer is  affirmative. To see this, let \textbf{A}, \textbf{B}, and \textbf{C} be three observables such that  [\textbf{A},\textbf{B}]=0 and  [\textbf{B},\textbf{C}]=0. If QC arises from the measurement statistics of these three observables, then the commutator [\textbf{A},\textbf{C}] must be {\em non-vanishing}, which is exactly a manifestation of quantum complementarity (see also Ref.~\cite{Clifton}). In fact, even if the QC is revealed by genuine POVM, a similar argument would show that QC must imply measurements that are not-jointly-measurable. Our proof thus gives a clear illustration of this close connection between complementarity and contextuality.
A natural line of research that stems from this observation is thus to establish this connection at a more quantitative level (e.g., in a more general framework): does the extent of complementarity also determine completely the extent that a system can be incompatible with noncontextuality and {\em vice versa} (see, e.g., the work that tries to answer this using the exclusivity principle~\cite{exclusivity})? Answers to all these questions would certainly lead to a better understanding of these peculiar features offered by quantum theory.

\vspace{5mm}
\noindent \textbf{Methods}

\noindent\textbf{Proof of Equation~(\ref{correspond}).} We need to compare the matrix elements
\begin{eqnarray}
\langle\Omega|(a_M)^{\nu_M}\cdots(a_2)^{\nu_2}(a_1)^{\nu_1}\sigma_1^{\hat n_1}\sigma_2^{\hat n_2}\cdots \sigma_M^{\hat n_M}\;\;\;\;\;\;\;\;\;\;\;\nonumber\\
\times(a_1^\dagger)^{\mu_1}(a_2^\dagger)^{\mu_2}\cdots (a_M^\dagger)^{\mu_M}|\Omega\rangle\label{element-QFT}
\end{eqnarray}
and
\begin{eqnarray}
\langle\nu_1\nu_2\cdots\nu_M|\hat\sigma_1^{\hat n_1}\otimes \hat\sigma_2^{\hat n_2}\otimes\cdots\otimes\hat\sigma_M^{\hat n_M}|\mu_1\mu_2\cdots\mu_M\rangle. \label{element-QM}
\end{eqnarray}
For the convenience to evaluate (\ref{element-QFT}), we have some useful relations
\begin{subequations}
\begin{eqnarray}
&\{2a_j^\dagger a_j-1,a_j^\dagger\}=0,&\\
&[2a_j^\dagger a_j-1,a_{j'}^\dagger]=0,{\rm \;for\;}j\neq j',&\\
&\{e^{i\phi_j}a_j+e^{-i\phi_j}a_j^\dagger,a_{j'}^\dagger\}=0, {\rm \;for\;}j\neq j'.&
\end{eqnarray}
\end{subequations}
Given these, one can readily verify
\begin{subequations}
\begin{eqnarray}
\sigma_1\sigma_2\cdots\sigma_{j-1}\sigma_j\cdots\sigma_M a_j^\dagger=\sigma_1\sigma_2\cdots\sigma_{j-1}\sigma_j a_j^\dagger\cdots\sigma_M,\\
a_j\sigma_1\sigma_2\cdots\sigma_{j-1}\sigma_j\cdots\sigma_M=\bar\sigma_1\bar\sigma_2\cdots\bar\sigma_{j-1} a_j\sigma_j\cdots\sigma_M,
\end{eqnarray}
\end{subequations}
where $\sigma_j\equiv\sigma_j^{\hat n_j}$ with $\hat n_j$ omitted without confusion, and $\bar\sigma_j\equiv\sigma_j(\bar\theta_j,\phi_j)$ with $\bar\theta_j=-\theta_j$.

In this way, the operator between $\langle\Omega|$ and $|\Omega\rangle$ in (\ref{element-QFT}) can be manipulated into a \emph{standard} form
\begin{eqnarray}
(-1)^\eta \biggr[(a_1)^{\nu_1}\tilde\sigma_1(a_1^\dagger)^{\mu_1}\biggr]\biggr[(a_2)^{\nu_2}\tilde\sigma_2(a_2^\dagger)^{\mu_2}\biggr]
\;\;\;\;\;\;\;\;\;\;\;\nonumber\\
\cdots\times\biggr[(a_M)^{\nu_M}\tilde\sigma_M(a_M^\dagger)^{\mu_M}\biggr],
\end{eqnarray}
where $\tilde\sigma_j=\sigma_j(\tilde\theta_j,\phi_j)$, $\tilde\theta_j=(-1)^{\xi_j}\theta_j$, indicating that $\theta_j$ could change its sign, depending on the times of swapping $a_j$ and $\sigma_{j'}$ ($j>j'$).
The $(-1)^\eta $ sign depends on various aspects: (i) the number of times $a_{j}$ and $a_{j'}$ ($j<j'$) are swapped, and (ii) the times of swapping $a_j$ and $a_{j'}^\dagger$ ($j>j'$), before reaching the above standard form. Each swap in (i) and (ii) contributes a minus sign. Explicitly, we have
\begin{eqnarray}
\eta&=&(\mu_1+\nu_1)(\nu_2+\nu_3+\cdots+\nu_M)\nonumber\\
&&+(\mu_2+\nu_2)(\nu_3+\cdots+\nu_M)\nonumber\\
&&+\cdots+(\mu_{M-1}+\nu_{M-1})\nu_M\nonumber\\
&=& \sum_{s=1}^{M-1}\sum_{t=s+1}^M (\mu_s+\nu_s)\nu_t,
\end{eqnarray}
and
\begin{eqnarray}
\xi_j=\sum_{s=j+1}^M \nu_s, {\rm \;\;\;\;with\;}\xi_M\equiv0.
\end{eqnarray}
The next step is to evaluate quantities in the square bracket in the standard form. It is found that
\begin{eqnarray}
&&(a_j)^{\nu_j}\sigma_j(a_j^\dagger)^{\mu_j}\nonumber\\
&=&\biggr(\prod_{m=1}^{j-1}(1-2a_m^\dagger a_m)\biggr)\nonumber\\
&&\times\biggr(e^{i\phi_j}(a_j)^{\nu_j}a_j (a_j^\dagger)^{\mu_j}
+e^{-i\phi_j}(a_j)^{\nu_j}a_j^\dagger(a_j^\dagger)^{\mu_j}\biggr)\sin\tilde\theta_j\nonumber\\
&&+\biggr(2(a_j)^{\nu_j}a_j^\dagger a_j(a_j^\dagger)^{\mu_j}-(a_j)^{\nu_j}(a_j^\dagger)^{\mu_j}\biggr)\cos\tilde\theta_j.
\end{eqnarray}
The quantity in the first bracket must be 1, due to $\langle\Omega|a_m^\dagger a_m|\Omega\rangle=0$. Note that a single $a_j$ or $a_j^\dagger$ does not survive between $\langle\Omega|$ and $|\Omega\rangle$, and that only $a_j a_j^\dagger$ will contribute. As a result, we have
\begin{eqnarray}
(a_j)^{\nu_j}\sigma_j(a_j^\dagger)^{\mu_j}=&-\cos\tilde\theta_j&{\rm\;for\;}\mu_j=0,\nu_j=0,\nonumber\\
=&\sin\tilde\theta_j e^{-i\phi_j}&{\rm\;for\;}\mu_j=0,\nu_j=1,\nonumber\\
=&\sin\tilde\theta_j e^{i\phi_j}&{\rm\;for\;}\mu_j=1,\nu_j=0,\nonumber\\
=&\cos\tilde\theta_j&{\rm\;for\;}\mu_j=1,\nu_j=1.
\end{eqnarray}
Hence, a product of such terms, together with signs determined by $\eta$ and $\xi_j$, constitutes (\ref{element-QFT}). On the other hand, (\ref{element-QM}) can be calculated explicitly, since it is factorizable. Then a direct comparison shows that (\ref{element-QFT}) and (\ref{element-QM}) are the same. This ends the proof.

Note that the above proof is for full correlations like Eqs. (\ref{element-QFT}) and (\ref{element-QM}). For partial correlations where the number of Pauli operators is less than $M$ (e.g., $\langle\nu_1\nu_2\cdots\nu_M|\hat\sigma_1^{\hat n_1}\otimes \hat\sigma_2^{\hat n_2}\otimes\cdots\otimes\hat\sigma_{M-1}^{\hat n_{M-1}}\otimes\openone|\mu_1\mu_2\cdots\mu_M\rangle$), the proof of correspondence is quite similar.

{\bf Acknowledgements}

We thank A. Cabello for valuable comments on an earlier version of this manuscript and to Frank Steinhoff for bringing Ref.~\cite{Clifton} to our attention. J.L.C. is supported by the National Basic Research Program (973
Program) of China under Grant No.\ 2012CB921900 and the NSF of China
(Grant Nos. 11175089 and 11475089). Y. C. L is supported by the Swiss NCCR QSIT and the ERC grant
258932. This work is also partly supported by the National Research Foundation
and the Ministry of Education, Singapore.

{\bf Author Contributions}

H.Y.S. devised the initial version of the result. H.Y.S., J.L.C., and Y.C.L. contributed to improvements in the result and the writing of the manuscript. All authors reviewed the manuscript.

{\bf Additional Information}

\textbf{Competing financial interests:} The authors declare no
competing financial interests.

\textbf{Reprints and permission} information is available at
www.nature.com/reprints.


\begin{thebibliography}{99}
\bibitem{EPR}
Einstein, A., Podolsky, B. \& Rosen, N. Can quantum-mechanical description of physical reality be considered complete? \emph{Phys. Rev.} {\bf 47}, 777 (1935).

\bibitem{Bohm1}
Bohm, D. A suggested interpretation of the quantum theory in terms of "hidden" variables. I. \emph{Phys. Rev. }{\bf 85}, 166 (1952).

\bibitem{Bohm2}
Bohm, D. A suggested interpretation of the quantum theory in terms of "hidden" variables. II. \emph{Phys. Rev.} {\bf 85}, 180 (1952).

\bibitem{Bell:RMP}
Bell, J. S.
On the problem of hidden variables in quantum mechanics.
\emph{Rev. Mod. Phys.} \textbf{38}, 447-452 (1966).

\bibitem{KS}
Kochen, S. \& Specker, E. P.
The problem of hidden variables in quantum  mechanics.
\emph{J. Math. Mech.} \textbf{17}, 59-87 (1967).

\bibitem{Peres}
Peres, A. \emph{Quantum Theory: Concepts and Methods} (Kluwer Academic Publishers, New York, Boston, Dordrecht, London, Moscow, 2002).

\bibitem{Spekkens:PRA:71} 
Spekkens, R. W. Contextuality for preparations, transformations, and unsharp measurements. \emph{Phys. Rev. A} {\bf 71}, 052108 (2005).

\bibitem{Bell64}
Bell, J. S. On the einstein podolsky rosen paradox. \emph{Physics} (NY) {\bf 1}, 195-200 (1964).

\bibitem{Brunner:RMP}
Brunner, N., Cavalcanti, D., Pironio, S., Scarani, V. \& Wehner, S. Bell nonlocality. \emph{Rev. Mod. Phys.} {\bf 86}, 419 (2014).

\bibitem{Foundations1}
Clauser, J. F. \& Shimony, A. Bell's theorem: Experimental tests and implications. \emph{Rep. Prog. Phys.} {\bf 41}, 1881 (1978).

\bibitem{Foundations2}
Mermin, N. D. Hidden variables and the two theorems of John Bell. \emph{Rev. Mod. Phys.} \textbf{65}, 803 (1993).

\bibitem{Foundations3}
Mermin, N. D. What is quantum mechanics trying to tell us? \emph{Am. J. Phys.} {\bf 66}, 753 (1998).

\bibitem{Foundations4}
Bell, J. S. {\em Speakable and unspeakable in quantum mechanics}
(Cambridge University Press, Cambridge, 2004).

\bibitem{Foundations5}
Colbeck, R. \& Renner, R. No extension of quantum theory can have improved predictive power. \emph{Nat. Commun.} {\bf 2}, 411; DOI:10.1038/ncomms1416 (2011).

\bibitem{Foundations6}
Bancal, J.-D. \emph{et al.}
Quantum non-locality based on finite-speed causal influences leads to superluminal signalling. \emph{Nat. Phys.} {\bf 8}, 867 (2012).

\bibitem{Foundations7}
Wood, C. J. \& Spekkens, R. W. The lesson of causal discovery algorithms for quantum correlations: Causal explanations of Bell-inequality violations require fine-tuning. \emph{arXiv}:1208.4119 (2012) (Date of access:05/12/2014).

\bibitem{Foundations8}
Leifer, M. S. \& Maroney, O. J. E. Maximally epistemic interpretations of the quantum state and contextuality. \emph{Phys. Rev. Lett.} {\bf 110}, 120401 (2013).

\bibitem{Spekkens:1312.3667}
Spekkens, R. W. The status of determinism in proofs of the impossibility of a noncontextual model of quantum theory. \emph{Found. Phys.} \textbf{44}, 1125 (2014).

\bibitem{Foundations10}
Vona, N. \& Liang, Y.-C. Bell's theorem, accountability and nonlocality. \emph{J. Phys. A: Math. Theor.} \textbf{47}, 424026 (2014).

\bibitem{Liang}
Liang, Y.-C., Spekkens, R. W. \& Wiseman, H. M. Specker's parable of the overprotective seer: A road to contextuality,
nonlocality and complementarity. \emph{Phys. Rep.} \textbf{506}, 1 (2011).

\bibitem{Scarani:DIQIP}
Scarani, V. The device-independent outlook on quantum physics. \emph{Acta Phys. Slovaca} {\bf 62}, 347-409 (2012).

\bibitem{Howard:Nature:2014}
Howard, M., Wallman, J., Veitch, V. \& Emerson, J. Contextuality supplies the `magic' for quantum computation. \emph{Nature} {\bf 510}, 351 (2014).

\bibitem{Nielsen}
Nielsen, M. A. \& Chuang, I. {\em Quantum Computation and Quantum Information Theory} (Cambridge University Press, 2011).

\bibitem{KCBS}
Klyachko, A. A., Can, M. A., Binicio\v{g}lu, S. \& Shumovsky, A. S. Simple test for hidden variables in spin-1 systems. \emph{Phys. Rev. Lett.} {\bf 101}, 020403 (2008).

\bibitem{Cabello:1998} 
Cabello, A. \& Garcia-Alcaine, G. Proposed experimental tests of the Bell-Kochen-Specker theorem. \emph{Phys. Rev. Lett.} \textbf{80}, 1797 (1998).

\bibitem{Simon:PRL} 
Simon, C. Brukner, C. \& Zeilinger, A. Hidden Variable
Theorems for Real Experiments. \emph{Phys. Rev. Lett.} \textbf{86}, 4427 (2001).

\bibitem{Larsson} 
Larsson, J.-A. A Kochen-Specker Inequality, \emph{Europhys. Lett.} \textbf{58}, 799 (2002).

\bibitem{Cabello08}
Cabello, A. Experimentally testable state-independent quantum contextuality. \emph{Phys. Rev. Lett.} {\bf 101}, 210401 (2008).

\bibitem{Cabello09}
Badzikag, P., Bengtsson, I.,
Cabello, A. \& Pitowsky, I. Universality of state-independent violation of correlation inequalities for noncontextual theories. \emph{Phys. Rev. Lett.} \textbf{103}, 050401
(2009).

\bibitem{spin-1/2}
Chen, J. L. \emph{et al.}
Quantum contextuality for a relativistic spin-1/2 particle.
\emph{Phys. Rev. A} \textbf{87}, 022109 (2013).

\bibitem{Araujo}
Ara\'ujo, M. {\em et al.}
All noncontextuality inequalities for the n-cycle scenario.
\emph{Phys. Rev. A} \textbf{88}, 022118 (2013).

\bibitem{NCI}
Yu, X.-D. \& Tong, D. M. Coexistence of Kochen-Specker inequalities and noncontextuality inequalities. \emph{Phys. Rev. A} {\bf 89}, 010101 (2014).

\bibitem{KCBS-exp}
Lapkiewicz, R. \emph{et al.}
Experimental non-classicality of an indivisible quantum system.
\emph{Nature} \textbf{474}, 490-493 (2011).

\bibitem{Blatt09}
Kirchmair, G. \emph{et al.}
State-independent experimental test of quantum contextuality.
\emph{Nature} \textbf{460}, 494 (2009).

\bibitem{exp1}
Amselem, E.,
R{\aa}dmark, M., Bourennane, M. \& Cabello, A. State-independent quantum contextuality with single photons. \emph{Phys. Rev. Lett.}
\textbf{103}, 160405 (2009).

\bibitem{exp2}
Moussa, O., Ryan, C. A., Cory, D. G. \&
Laflamme, R. Testing contextuality on quantum ensembles with one clean qubit. \emph{Phys. Rev. Lett.} \textbf{104}, 160501 (2010).

\bibitem{exp3}
Zhang, X. \emph{et al.}
State-independent experimental test of quantum contextuality with a single trapped ion. \emph{Phys. Rev. Lett.} \textbf{110}, 070401 (2013).

\bibitem{Meyer} 
Meyer, D. A. Finite Precision Measurement Nullifies the Kochen-Specker Theorem. \emph{Phys. Rev. Lett.} \textbf{83}, 3751 (1999).

\bibitem{Kent} 
Kent, A. Noncontextual Hidden Variables and Physical Measurements
\emph{Phys. Rev. Lett.} \textbf{83}, 3755 (1999).

\bibitem{KentClifton} 
Clifton, R. \& Kent, A. Simulating quantum mechanics by non-contextual hidden variables. \emph{Proc. R. Sco. Lond. A} \textbf{456}, 2101 (2000).

\bibitem{Winter:JPA} 
Winter, A. What does an experimental test of quantum contextuality prove or disprove?
\emph{J. Phys. A: Math. Theor.} \textbf{47}, 424031 (2014).

\bibitem{Cabello:PRL:90}
Cabello, A. Kochen-Specker theorem for a single qubit using positive operator-valued measures. \emph{Phys. Rev. Lett.} {\bf 90}, 190401 (2003).

\bibitem{Busch:PRL} 
Busch, P. Quantum states and generalized observables: a simple proof of Gleason's theorem. {\em Phys. Rev. Lett.} \textbf{91}, 120403 (2003).

\bibitem{criticize}
Grudka, A. \& Kurzy\'nski, P. Is there contextuality for a single qubit? \emph{Phys. Rev. Lett.} \textbf{100}, 160401 (2008).

\bibitem{identical}
Cabello, A. \& Cunha, M. T.
State-independent contextuality with identical particles.
\emph{Phys. Rev. A} \textbf{87} 022126 (2013).

\bibitem{single}
Peskin, M. E. \& Schroeder, D. V. \emph{An Introduction to Quantum Field Theory} (Addison-Wesley, Reading, MA, 1995).

\bibitem{JW-trans}
Jordan, P. \& Wigner, E. \"{U}ber das Paulische \"{A}quivalenzverbot. \emph{Zeitschrift f\"{u}r Physik} \textbf{47}, 631-651 (1928).

\bibitem{CHSH}
Clauser, J. F., Horne, M. A., Shimony, A. \& Holt, R. A. Proposed experiment to test local hidden-variable theories. \emph{Phys. Rev. Lett.} {\bf 23}, 880 (1969).

\bibitem{Mermin}
Mermin, N. D.
Simple unified form for the major no-hidden-variables theorems.
 \emph{Phys. Rev. Lett.} \textbf{65}, 3373 (1990).

\bibitem{Mermin-Peres}
Peres, A. Incompatible results of quantum measurements.
\emph{ Phys. Lett. A} \textbf{151}, 107-108 (1990).

\bibitem{Hardy1}
Hardy, L. Nonlocality for two particles without inequalities for almost all entangled states. \emph{Phys. Rev. Lett.} \textbf{71}, 1665 (1993).

\bibitem{Hardy2}
Cereceda, J. L. Hardy's nonlocality for generalized n-particle GHZ states. \emph{Phys. Lett. A} \textbf{327}, 433 (2004).

\bibitem{Yu}
Yu, S., Chen, Q., Zhang, C., Lai, C. H. \& Oh, C. H. All entangled pure states violate a single Bell's inequality.
\emph{Phys. Rev. Lett.} \textbf{109}, 120402 (2012).

\bibitem{W-state1}
D\"ur, W., Vidal, G. \& Cirac, J. I.
Three qubits can be entangled in two inequivalent ways.
\emph{Phys. Rev. A} \textbf{62}, 062314 (2000).

\bibitem{planat12}
Planat, M. R. P. Quantum states arising from the Pauli groups, symmetries and paradoxes. \emph{arXiv}:1209.5176 (2012) (Date of access:24/09/2012).

\bibitem{Dicke}
Dicke, R. Coherence in spontaneous radiation processes. \emph{Phys. Rev.} {\bf 93} 99 (1954).

\bibitem{Clifton}
Clifton, R. Complementarity between position and momentum as a consequence of Kochen-Specker arguments. \emph{Phys. Lett. A} {\bf 271}, 1 (2000).

\bibitem{exclusivity}
Cabello, A. Simple explanation of the quantum violation of a fundamental inequality. \emph{Phys. Rev. Lett.} \textbf{110}, 060402 (2013).


\end{thebibliography}
\end{document}